\documentclass[aps,prl,twocolumn,superscriptaddress,longbibliography,]{revtex4-1}

\usepackage{amssymb}
\usepackage{graphicx}
\usepackage{amsmath}
\usepackage{hyperref}
\usepackage{xspace}
\usepackage{xcolor}
\AtBeginDocument{\usepackage{booktabs}}

\pdfoutput=1

\newcommand{\MSbar}{\overline{\text{MS}}}
\newcommand{\sss}{\scriptscriptstyle}

\newcommand{\as}{\alpha_s}

\newcommand{\Lum}{\mathcal{L}}

\newcommand{\mh}{m_{\sss\rm H}}
\newcommand{\mt}{m_{\sss\rm t}}
\newcommand{\mbottom}{m_{\sss\rm b}}
\newcommand{\mcharm}{m_{\sss\rm c}}

\newcommand{\muf}{\mu_{\sss\rm F}}
\newcommand{\mur}{\mu_{\sss\rm R}}

\let\originalleft\left
\let\originalright\right
\renewcommand{\left}{\mathopen{}\mathclose\bgroup\originalleft}
\renewcommand{\right}{\aftergroup\egroup\originalright}

\def\beq{\begin{equation}}  
\def\eeq{\end{equation}}
\def\({\left(}
\def\){\right)}
\def\[{\left[}
\def\]{\right]}

\begin{document}

\preprint{}

\title{Double resummation for Higgs production}


\author{Marco Bonvini}
\email{marco.bonvini@roma1.infn.it}
\affiliation{INFN, Sezione di Roma 1, Piazzale Aldo Moro~5, 00185 Roma, Italy}

\author{Simone Marzani}
\email{simone.marzani@ge.infn.it}
\affiliation{Dipartimento di Fisica, Universit\`a di  Genova and INFN, Sezione di Genova, Via Dodecaneso 33, I-16146 Genova, Italy}

\date{\today}

\date{\today}

\begin{abstract}
  We present the first double-resummed prediction of the inclusive
  cross section for the main Higgs production channel in proton-proton
  collisions, namely gluon fusion.
  Our calculation incorporates to all orders in perturbation theory
  two distinct towers of logarithmic corrections which are enhanced,
  respectively, at threshold, i.e.\ large~$x$, and in the high-energy
  limit, i.e.\ small~$x$. Large-$x$ logarithms are resummed to
  next-to-next-to-next-to-leading logarithmic accuracy, while
  small-$x$ ones to leading logarithmic accuracy. The double-resummed
  cross section is furthermore matched to the state-of-the-art fixed-order
  prediction at next-to-next-to-next-to-leading accuracy.
  We find that double resummation corrects the Higgs production rate
  by 2\% at the currently explored center-of-mass energy of 13~TeV and
  its impact reaches 10\% at future circular colliders at 100~TeV. 
\end{abstract}

%\pacs{}
%\keywords{}

\maketitle

\begin{figure*}[t]
  \includegraphics[width=0.49\textwidth,page=1]{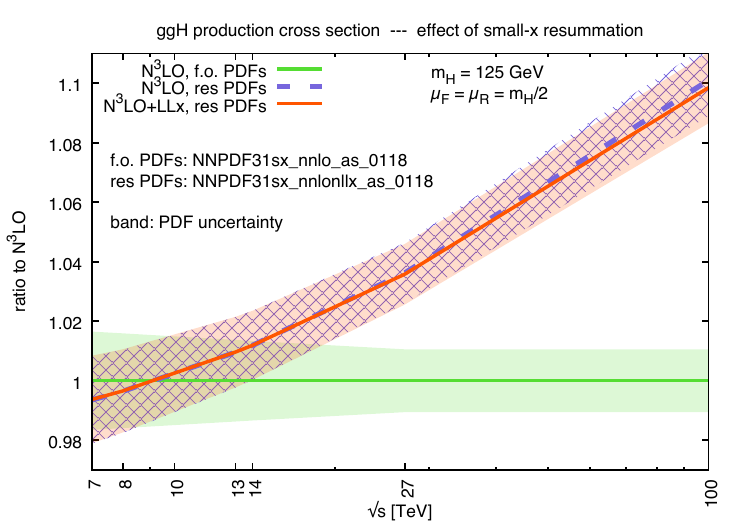}
  \includegraphics[width=0.49\textwidth,page=2]{images/ggH_Kfactor_summary.pdf}
  \caption{All-order effects on the Higgs cross section computed at N$^3$LO, as a function of $\sqrt{s}$. The plot of the left shows the impact of small-$x$ resummation, while the one of the right of large-$x$ resummation. The bands represent PDF uncertainties.}\label{fig:shat}
\end{figure*}

The major achievement of the first run of the CERN Large Hadron
Collider (LHC) was the discovery of the Higgs
boson~\cite{Aad:2012tfa,Chatrchyan:2012ufa}, thus confirming the
Brout-Englert-Higgs mechanism~\cite{Englert:1964et, Higgs:1964ia,
  Higgs:1964pj, Higgs:1966ev} for the electroweak symmetry breaking.
The current and future runs of the LHC are rightly considered the Higgs
precision era. 
The ATLAS and CMS collaborations are continuously producing experimental analyses of ever increasing
sophistication, which allow for a more detailed inspection of the Higgs
sector. 
Examples include measurements of fiducial cross sections in
different decay channels, as well as kinematic distributions of the
Higgs
boson~\cite{Aaboud:2017oem,Aaboud:2017vzb,Khachatryan:2016vau,Aad:2016lvc,Aad:2015gba,Aad:2015lha,Sirunyan:2018qio,Sirunyan:2017exp,Khachatryan:2016vnn,Khachatryan:2015yvw,Khachatryan:2015rxa}.
In order to perform meaningful comparisons, it is imperative for the
theoretical physics community to deliver calculations with
uncertainties that match in magnitude those quoted by the experimental
collaborations.
Therefore, it does not come as a surprise that perturbative
calculations in QCD have reached the astonishing
next-to-next-to-next-to-leading order (N$^3$LO) precision for the two
most important Higgs production channels in proton-proton collisions:
gluon fusion (GF)~\cite{Anastasiou:2014vaa, Anastasiou:2014lda,
  Anastasiou:2015ema,Anastasiou:2016cez,Mistlberger:2018etf,Li:2014afw,Li:2014bfa,Anzai:2015wma} and vector-boson
fusion~\cite{Dreyer:2016oyx}.

The calculation of the inclusive GF cross section is particularly
challenging in QCD because, even at its lowest order, it proceeds via
a massive quark loop~\cite{Georgi:1977gs}.
Furthermore, this process is characterized by large perturbative
QCD corrections.  The NLO contribution (originally computed
in~\cite{Dawson:1990zj,Djouadi:1991tka} in an effective field theory (EFT) approximation
where the top-quark mass is considered much larger than any other scale,
and in~\cite{Spira:1995rr} for general quark mass running in the loop) is
as large as the leading order, and the NNLO corrections (computed
in~\cite{Harlander:2002wh,Anastasiou:2002yz,Ravindran:2003um} in the
EFT and with finite top-mass corrections
in~\cite{Harlander:2009bw,Harlander:2009mq,Harlander:2009my,Pak:2009bx,Pak:2009dg})
are about half as large as the LO.
The aforementioned EFT three-loop calculation of~\cite{Anastasiou:2014vaa,
  Anastasiou:2014lda, Anastasiou:2015ema,Anastasiou:2016cez,Mistlberger:2018etf} finally
shows perturbative convergence, with small theoretical uncertainties,
as estimated by varying the arbitrary scales of the perturbative
calculation.

A complementary approach to include higher-order corrections can be
pursued using all-order techniques, which rely upon factorization
properties of QCD matrix elements, and phase-space, in particular
kinematic limits.
In the case of the inclusive GF cross section, one of the most studied and
well-developed all-order techniques is threshold (or large~$x$)
resummation~\cite{Catani:1996yz,deFlorian:2012yg,Ahrens:2008nc}, which
has been pushed to next-to-next-to-next-to-leading logarithmic
(N$^3$LL) accuracy~\cite{Moch:2005ba,Moch:2005ky,Laenen:2005uz,Anastasiou:2014vaa,Bonvini:2014joa,Catani:2014uta,Bonvini:2014tea,Schmidt:2015cea},
%~\footnote{To be precise, all contributions that are relevant for N$^3$LL are
%known~\cite{Moch:2005ba,Moch:2005ky,Laenen:2005uz,Anastasiou:2014vaa},
%with the exception of the four-loop cusp anomalous dimension
%(see~\cite{Boels:2017skl,Grozin:2017css} for recent progress), which is
%thought to have a negligible impact.}
and leads to a significantly improved convergence of the perturbative expansion~\cite{Bonvini:2016frm}.
In the opposite kinematic limit, i.e.\ at high energy (or small~$x$),
resummation~\cite{Fadin:1975cb,Lipatov:1976zz,Kuraev:1976ge,Kuraev:1977fs,Balitsky:1978ic}
is also possible.
In particular, small-$x$ resummation of splitting functions~\cite{Catani:1993rn}, which govern the evolution of the parton distribution functions (PDFs), was obtained in~\cite{Ball:1997vf,Ball:1999sh,Altarelli:1999vw,Altarelli:2000mh,Altarelli:2001ji,Altarelli:2003hk,Altarelli:2005ni,Ball:2007ra,Altarelli:2008xp,Altarelli:2008aj} (see also~\cite{Salam:1998tj,Ciafaloni:1998iv,Ciafaloni:1999yw,Ciafaloni:1999au,Ciafaloni:2000cb,Ciafaloni:2002xf,Ciafaloni:2003rd,Ciafaloni:2003kd,Ciafaloni:2005cg,Ciafaloni:2006yk,Ciafaloni:2007gf,Thorne:1999sg,Thorne:1999rb,Thorne:2001nr,White:2006yh}).
Furthermore, high-energy resummation of partonic coefficient functions can be achieved in the framework of
$k_t$-factorization~\cite{Catani:1990eg,Catani:1994sq}, and improved
with the inclusion of subleading (but important) running coupling
corrections~\cite{Ball:2007ra}.
All-order calculations in this framework have been exploited to assess
and improve the aforementioned EFT, which is known to fail in the
high-energy limit~\cite{Hautmann:2002tu,
  Marzani:2008az,Pasechnik:2006du}.
However, thus far, investigations in the high-energy regime were
performed at fixed-order, i.e.\ by expanding out the all-order
results.
All-order phenomenological results at small~$x$ will be presented in this work for the first time.

In this letter, we combine the above all-order approaches, together
with the state-of-the-art fixed-order calculation, to obtain the most
accurate prediction for Higgs production at the LHC.
To our knowledge, it is the first time that double (large- and small-$x$) resummation is achieved.
This breakthrough is possible because of two distinct advancements in
the field. On the one hand, a general framework to combine the two
resummations has been developed in~\cite{Ball:2013bra} and
implemented in public codes
\texttt{TROLL}~\cite{Bonvini:2014joa,Bonvini:2016frm} and
\texttt{HELL}~\cite{Bonvini:2016wki,Bonvini:2017ogt} so that numerical
results can be easily obtained. On the other hand, recently, all-order
calculations have been considered in the context of PDF determination, both at
large-$x$~\cite{Bonvini:2015ira} and at
small-$x$~\cite{Ball:2017otu,Abdolmaleki:2018jln}. This opens up the possibility of
achieving fully consistent resummed results.
While we presently concentrate on the Higgs production cross section,
our technique is fully general and can be applied to other important
processes, such as the Drell-Yan process or heavy-quark production. We
leave further phenomenological analyses to future work.

\begin{figure*}[t]
  \includegraphics[width=0.32\textwidth,page=3]{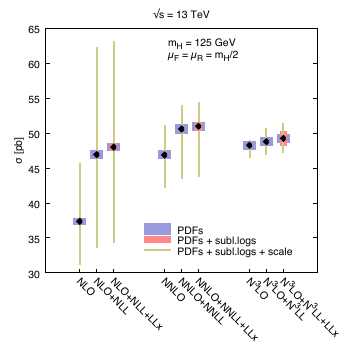}
  \includegraphics[width=0.32\textwidth,page=5]{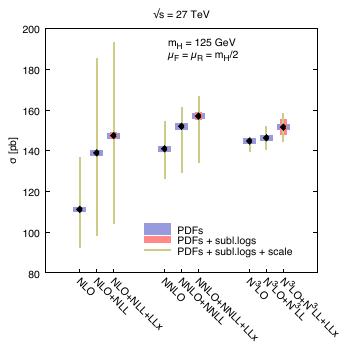}
  \includegraphics[width=0.32\textwidth,page=6]{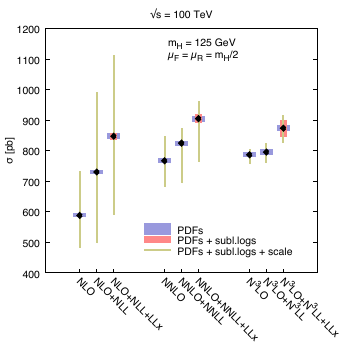}
  \caption{Perturbative progression of the Higgs inclusive cross section in different approximations:
    fixed-order, threshold resummation and double resummation, at three representative values of the collision energy.}
  \label{fig:progression}
\end{figure*}

Let us start our discussion by introducing the factorized Higgs production cross section
\begin{align}\label{eq:xs}
\sigma(\tau,\mh^2) &= \tau \sigma_0\(\mh^2,\as(\mur^2)\)  \\ & \times
\sum_{ij}\int_\tau^1 \tfrac{dx}{x}\,\Lum_{ij}\(\tfrac{\tau}{x},\muf^2\)
 C_{ij}\(x,\as(\mur^2),\tfrac{\mh^2}{\muf^2},\tfrac{\mh^2}{\mur^2}\),\nonumber
\end{align}
where $\sigma_0$ is the lowest-order partonic cross section, $\Lum_{ij}$ are parton
luminosities (convolutions of PDFs), $C_{ij}$ are the perturbative partonic coefficient functions,
$\tau=\mh^2/s$ is the squared ratio between
the Higgs mass and the collider center-of-mass energy,
and the sum runs over all parton flavors.
Henceforth, we suppress the dependence on renormalization and
factorization scales $\mur,\muf$. Moreover, because the
Higgs couples to the gluon via a heavy-flavor loop, \eqref{eq:xs} also implicitly
depends on any heavy virtual particle mass.

The general method to consistently combine large- and small-$x$
resummation of partonic coefficient functions $C_{ij}(x,\as)$ was
developed in~\cite{Ball:2013bra}.
The basic principle is the definition of each resummation such that they do not
interfere with each other.
This statement can be made more precise by considering Mellin ($N$) moments of~(\ref{eq:xs}).
The key observation is that while in momentum ($x$) space coefficient functions are distributions, their Mellin moments are analytic functions of the complex variable $N$ and therefore, they are (in principle) fully determined by the knowledge of their singularities. 
Thus, high-energy and threshold resummations are consistently combined if they mutually respect their singularity structure.
In~\cite{Ball:2013bra}, where an approximate N$^3$LO result for $C_{ij}$ was obtained by expanding both resummations to $\mathcal{O}(\as^3)$,
the definition of the large-$x$ logarithms from threshold resummation was improved in order to satisfy the desired behavior,
and later this improvement was extended to all orders in~\cite{Bonvini:2014joa}, leading to the so-called $\psi$-soft resummation scheme.
Thanks to these developments, double-resummed partonic coefficient functions can be simply written as the sum of three terms~\footnote{The fixed-order and small-$x$ contributions are written in this form, while the large-$x$ one is written in $N$ space and its Mellin inversion is performed together with the parton luminosity using the so-called minimal prescription~\cite{Catani:1996yz}.}
\begin{equation} \label{eq:master}
C_{ij}\(x,\as\)= C_{ij}^\text{fo}\(x,\as\)+ \Delta C_{ij}^\text{l$x$}\(x,\as\)+\Delta C_{ij}^\text{s$x$}\(x,\as\),
\end{equation}
where the first term is the fixed-order calculation, the second one is
the threshold-resummed $\psi$-soft contribution minus its expansion (to avoid
double counting with the fixed-order), and the third one is the
resummation of small-$x$ contributions, again minus its
expansion. Note that not all partonic channels contribute to all terms
in \eqref{eq:master}. For instance, the $qg$ contribution is
power-suppressed at threshold but it does exhibit logarithmic enhancement at small~$x$.

Our result brings together the highest possible accuracy
in all three contributions. The fixed-order piece is
N$^3$LO~\cite{Anastasiou:2014vaa, Anastasiou:2014lda,
  Anastasiou:2015ema,Anastasiou:2016cez,Mistlberger:2018etf},
supplemented with the correct small-$x$ behavior, as implemented in the public
code {\tt ggHiggs}~\cite{Ball:2013bra,Bonvini:2014jma,Bonvini:2016frm}.
Threshold-enhanced contributions are accounted for to
next-to-next-to-next-to-leading logarithmic accuracy (N$^3$LL) in the
$\psi$-soft scheme, as implemented in the public code
\texttt{TROLL}~\cite{Bonvini:2014joa,Bonvini:2016frm}.
Finally, for high-energy resummation we consider the resummation of
the leading non-vanishing tower of logarithms (here LL$x$) to the
coefficient functions~\cite{Marzani:2008az,Ball:2007ra}, which we
have now implemented in the code~\texttt{HELL}~\cite{Bonvini:2016wki,Bonvini:2017ogt}.
The technical details of the implementation will be presented
elsewhere~\cite{Bonvini:2018iwt}.
Additionally, on top of scale variations,
subleading terms can be varied in both resummed contributions,
thus allowing for the estimate of the uncertainty from missing higher orders
and from the matching procedure.
Our calculation keeps finite top-mass effects where possible. In
particular, in the fixed-order part they are included up to NNLO and
in the threshold-resummed one up to NNLL. Furthermore, the small-$x$
contribution, both at fixed order and to all orders, must be computed with finite top-mass, essentially
because the limits $x\to 0$ and $\mt \to \infty$ do not commute.
We will discuss further corrections associated with the masses of bottom and charm quarks when presenting our final results.

\begin{table*}[t]
 \centering
 \begin{tabular}{r c ccc c ccccc}
   & \multicolumn{5}{c}{\texttt{NNPDF31sx\_nnlo\_as\_0118}} & \multicolumn{5}{c}{\texttt{NNPDF31sx\_nnlonllx\_as\_0118}} \\[1ex]
   \multicolumn{1}{c}{$\sqrt{s}$} &\hspace{2em}
   & $\sigma_\text{N$^3$LO}$ & $\delta_{\rm scale}$ & $\delta_{\rm PDFs}$ &\hspace{2em}
   & $\sigma_\text{N$^3$LO+N$^3$LL+LL$x$}$ & $\delta_{\rm scale}^{\rm 42var}$ & $\delta_{\rm PDFs}$ & $\delta_{\rm subl.logs}$ & $\Delta\sigma_{b,c}$  \\
\midrule
   7~TeV && $ 16.76$~pb & ${}_{ -3.7}^{+ 0.7}\%$ & $\pm 1.7\%$ && $ 16.83$~pb & ${}_{ -3.6}^{+ 4.2}\%$ & $\pm 1.5\%$ & $\pm 1.3\%$ & $ -1.01$~pb \\[1ex]
   8~TeV && $ 21.32$~pb & ${}_{ -3.7}^{+ 0.7}\%$ & $\pm 1.6\%$ && $ 21.47$~pb & ${}_{ -3.6}^{+ 4.1}\%$ & $\pm 1.4\%$ & $\pm 1.4\%$ & $ -1.26$~pb \\[1ex]
  13~TeV && $ 48.28$~pb & ${}_{ -3.7}^{+ 0.9}\%$ & $\pm 1.4\%$ && $ 49.26$~pb & ${}_{ -3.8}^{+ 4.0}\%$ & $\pm 1.2\%$ & $\pm 1.8\%$ & $ -2.66$~pb \\[1ex]
  14~TeV && $ 54.32$~pb & ${}_{ -3.7}^{+ 0.9}\%$ & $\pm 1.3\%$ && $ 55.56$~pb & ${}_{ -3.8}^{+ 4.0}\%$ & $\pm 1.2\%$ & $\pm 1.9\%$ & $ -2.96$~pb \\[1ex]
  27~TeV && $ 144.7$~pb & ${}_{ -3.7}^{+ 0.9}\%$ & $\pm 1.1\%$ && $ 151.6$~pb & ${}_{ -4.0}^{+ 4.0}\%$ & $\pm 1.0\%$ & $\pm 2.3\%$ & $  -7.2$~pb \\[1ex]
 100~TeV && $ 786.7$~pb & ${}_{ -3.8}^{+ 1.9}\%$ & $\pm 1.1\%$ && $ 873.9$~pb & ${}_{ -4.3}^{+ 4.0}\%$ & $\pm 1.2\%$ & $\pm 3.0\%$ & $ -32.0$~pb
 \end{tabular}
 \caption{Values of the N$^3$LO and N$^3$LO+N$^3$LL+LL$x$ GF cross section
   for selected values of the collider energy and Higgs mass $\mh=125$~GeV.
   We use the NNPDF31sx PDFs with $\as(m_Z^2)=0.118$, $\mt=173$~GeV, $\mbottom=4.92$~GeV and $\mcharm=1.51$~GeV.}
 \label{tab:xsec1}
\end{table*}

Having determined the resummation of the partonic coefficient
functions, we now discuss the role of the parton luminosities
$\Lum_{ij}$ that enter~\eqref{eq:xs}.
Ideally, we would like to use PDFs that have been fitted using a
double-resummed theory. However, this is clearly not possible. Indeed,
this is the first study that aims to combine threshold and high-energy
resummation, so a PDF fit with this theory will only appear in the
future.
Therefore, we have to find an acceptable compromise. Within the NNPDF
framework~\cite{Ball:2017nwa}, PDFs with threshold resummation were obtained
in~\cite{Bonvini:2015ira}, while small-$x$ resummation was considered
in~\cite{Ball:2017otu}. We note that the inclusion of the latter was a 
challenging enterprise because small-$x$ logarithms appear both in
coefficient functions and PDF evolution, while in the $\MSbar$ scheme
large-$x$ resummation only affects coefficient
functions~\cite{Korchemsky:1988si,Albino:2000cp}.
In order to make an informed decision, we separately consider in
Fig.~\ref{fig:shat} the impact of small-$x$ resummation (on the left)
and large-$x$ resummation (on the right) on the GF cross section, as a
function of the center-of-mass energy of the colliding protons.

Let us start by illustrating the situation concerning small-$x$ resummation (left-hand plot).
The plot shows the ratio of resummed results to the fixed order one, computed at N$^3$LO with the fixed-order
NNLO set of~\cite{Ball:2017otu}. We include resummation in two
steps. First (dashed blue), we compute the N$^3$LO cross section
using the ``resummed PDFs'' of~\cite{Ball:2017otu}, i.e.\ those fitted
including resummation and evolving with NNLO+NLL$x$ theory.  Then
(solid red), we add the LL$x$ resummation to the Higgs coefficient
functions, which provides the consistent resummed result.  In all
cases, the bands correspond to PDF uncertainties.
The plot clearly shows that small-$x$ resummation has a modest effect at
current LHC energies, but its impact grows substantially with the
energy, reaching the 10\% level at 100~TeV, heralding the fact that electroweak physics at 100~TeV is small-$x$ physics. 
The plot also shows that the bulk of the effect comes from the resummed PDFs and their resummed
evolution, while small-$x$ resummation of the Higgs coefficient functions is only a small correction.
This perhaps surprising result can be understood by noting that, while the high-energy behavior of the PDFs is essentially determined by deep-inelastic scattering data at small~$x$ and low $Q^2$, the Higgs cross section is characterized by a much higher value of $Q^2$, and it is dominated by soft emissions~\cite{Bonvini:2012an}.
Furthermore, the large discrepancy between resummed and NNLO PDFs at large $\sqrt{s}$ is a manifestation of the perturbative instability of the latter.
Indeed, as discussed at length in~\cite{Ball:2017otu}, resummed PDFs are close to the NLO ones, while the NNLO set significantly deviates at small~$x$.

The situation is rather different if we analyze large-$x$ resummation (right-hand plot).
Here we use the PDFs of~\cite{Bonvini:2015ira}, obtained with either NNLO and NNLO+NNLL theory,
which however suffer from a larger uncertainty compared to standard global fits because of the reduced dataset used in their determination.
In this case the impact of the resummation on the
N$^3$LO cross section is smaller and fairly constant in the whole
energy range considered here. The plot shows that about half of the
2\% effect originates from the resummation in the PDFs 
(dashed blue), which is however not significant due to the large PDF uncertainties,
and the other half by the resummation in the coefficient functions (solid red).

Therefore, by comparing the two plots in Fig.~\ref{fig:shat} we
conclude that, lacking double-resummed PDFs, the use of small-$x$
resummed PDFs is preferred for the fairly large energy range
considered here, because threshold-resummation effects in PDFs have a much
smaller impact on the Higgs cross section.
From the plots one may wonder whether double resummation of the
coefficient functions is at all needed for phenomenology. Certainly
its impact is numerically modest but we argue that its inclusion
brings significant advantages both at small- and large-$x$. Firstly,
it allows for a fully consistent treatment at small-$x$. Furthermore,
the inclusion of large-$x$ resummation, although being a small
correction to the N$^3$LO results, allows for a more robust estimate
of the theoretical uncertainty~\cite{Bonvini:2014joa,Bonvini:2016frm}.

We present double-resummed results for the Higgs cross section in
GF in Fig.~\ref{fig:progression}, where we show three plots at
representative center-of-mass energies of the colliding protons. 
We consider the current energy of the LHC, $\sqrt{s}=13$~TeV, and
two possible energies for future colliders, namely $\sqrt{s}=27$~TeV (HE-LHC) and
$\sqrt{s}=100$~TeV (FCC). 
We choose as central scale $\muf=\mur=\mh/2$.
Numerical results are presented in Tab.~\ref{tab:xsec1}, where we also report for completeness the correction $\Delta \sigma_{b,c}$ to the fixed-order calculation due to the presence of massive bottom and charm quarks running in the loop, following the recommendation of~\cite{deFlorian:2016spz}. 
Furthermore, electroweak corrections in the factorized approach, when included, amount to an extra 5\% increase~\cite{deFlorian:2016spz}.

Each plot shows the perturbative progression of the cross section as
obtained in different approximations: fixed-order, fixed-order and
threshold, fixed-order and double resummation.
We also show the three main contributions to the theoretical
uncertainty, namely PDFs, subleading logarithms at small-$x$ and scale
variation. The latter also includes an estimate of subleading
corrections at large~$x$, resulting in 42 variations, as detailed
in~\cite{Bonvini:2016frm}.
The uncertainty due to subleading logarithms at
small-$x$ has been determined by taking the envelope of two variants
of the coefficient-function resummation, which take as input resummed splitting functions either at
LL$x$ (to be precise, it is a modification of LL$x$ resummation which was
called LL$'$ in~\cite{Bonvini:2016wki,Bonvini:2017ogt}) or at
NLL$x$~\cite{Ball:2017otu,Bonvini:2018iwt}.
We note that the PDFs are, in principle, affected by analogous uncertainty, which however is not currently included in their determination. 
Thus, the overall small-$x$ uncertainty might be underestimated.
A qualitative assessment of this uncertainty was performed in~\cite{Ball:2017otu} and its impact on the Higgs cross section will be investigated in~\cite{Bonvini:2018iwt}.

We note that double resummation, mostly because of its threshold component, has a much more stable perturbative progression than its fixed-order counterpart: convergence is faster and uncertainty bands always cover the next perturbative order and shrink as higher orders are included~\cite{Bonvini:2016frm}. 
While double resummation is a small (2\%) correction to the N$^3$LO at current LHC energies, because of its small-$x$ component
its impact grows with $\sqrt{s}$, becoming 5\% at 27~TeV, before reaching approximately 10\% at 100~TeV.
Furthermore, we point out that a large contribution to the theoretical uncertainty
originates from unknown subleading logarithms at small~$x$.
As a consequence, our double-resummed prediction exhibits larger uncertainties than the N$^3$LO one.
On the one hand this highlights the importance of pushing the resummation of coefficient functions at
small~$x$ one order higher.
On the other hand, this also implies that the uncertainty from missing higher orders is likely underestimated in a purely fixed-order approach,
mostly due to the fact that PDF uncertainty does not fully account for it.
Thus, even at LHC energies where its impact is modest, double-resummation provides a more reliable estimate of the theoretical uncertainty affecting the Higgs cross section.

In this letter we have presented, for the first time, results in
perturbative QCD that supplement a fixed-order calculation with both
threshold and high-energy resummation. We have applied our
double-resummed framework to calculate the inclusive cross section for
Higgs production in gluon fusion. Our result features the state-of-the
art accuracy N$^3$LO+N$^3$LL+LL$x$ and crucially, it makes use of
recently determined resummed parton distributions.
The method presented here is rather general and it can be applied to a variety of processes currently studied at the LHC, such as electroweak-boson production or top-quark production. Furthermore, we anticipate that its generalization to differential distributions, such as rapidity and transverse momentum, is possible and we look forward to future work in this direction.

\begin{acknowledgments}
We thank R.~D.~Ball, S.~Forte, G.~Ridolfi, L.~Rottoli for many useful discussions.
The work of MB is supported by the Marie Sk\l{}odowska-Curie grant \mbox{HiPPiE@LHC.}
\end{acknowledgments}
%\newpage
\bibliography{higgs_biblio}
\end{document}